\documentclass[a4paper,11pt]{article}
\pdfoutput=1 
\usepackage{jcappub} 
\usepackage[T1]{fontenc} 
\usepackage{natbib}
\usepackage{graphics}
\usepackage{multirow}
\usepackage{amssymb}
\usepackage{amsmath}
\usepackage{mathrsfs}
\usepackage{xspace} 

\title{A Technique for Detection of PeV Neutrinos Using a Phased Radio Array}
\author[a,b]{A.~G.~Vieregg}
\author[b]{K.~Bechtol}
\author[c]{A.~Romero-Wolf}
\affiliation[a]{Department of Physics, Enrico Fermi Institute, University of Chicago, \\5640 S Ellis Avenue, Chicago, IL 60637, USA}
\affiliation[b]{Kavli Institute for Cosmological Physics, University of Chicago, \\933 E 56th Street, Chicago, IL 60637, USA}
\affiliation[c]{Jet Propulsion Laboratory, California Institute of Technology, \\4800 Oak Grove Drive, Pasadena, CA 91109, USA}

\emailAdd{avieregg@kicp.uchicago.edu}
\emailAdd{bechtol@kicp.uchicago.edu}
\emailAdd{andrew.romero-wolf@jpl.nasa.gov}

\abstract{
The detection of high energy neutrinos ($10^{15}-10^{20}$~eV) 
is an important step toward understanding the most energetic 
cosmic accelerators and would enable tests of fundamental physics at energy scales that cannot easily 
be achieved on Earth. In this energy range, there are two expected 
populations of neutrinos: the astrophysical flux observed with IceCube at lower energies ($\sim1$~PeV) and the 
predicted cosmogenic flux at higher energies ($\sim10^{18}$~eV).
Radio detector arrays such as RICE, ANITA, ARA, and ARIANNA exploit the Askaryan effect and 
the radio transparency of glacial ice, which together enable enormous 
volumes of ice to be monitored with sparse instrumentation.
We describe here the design for a phased radio array
that would lower the energy threshold of radio techniques to the PeV scale, 
allowing measurement of the astrophysical flux observed with IceCube over an extended energy range.
Meaningful energy overlap with optical Cherenkov telescopes could be used for energy calibration.
The phased radio array design would also provide more efficient 
coverage of the large effective volume required to discover cosmogenic neutrinos.
}
\begin{document}

\maketitle
\flushbottom

\section{Introduction} 
\label{sec:intro}

The IceCube experiment has recently detected neutrinos of astrophysical origin with energies 
up to $\sim$2~PeV~\cite{bigBird}.  This is a breakthrough in high energy astrophysics,
and represents the first observation of astrophysical neutrinos (except those seen from the sun and 
supernova 1987a~\cite{kamiokande1987,imb}).  
There are several candidate source classes that could make neutrinos up to 100 PeV, including active galactic nuclei, 
gamma-ray bursts, and starburst galaxies (for reviews, see~\cite{waxman_2013,anchordoqui_2014,murase_2014}).
In the coming years, we seek to identify the currently unknown sources of PeV astrophysical neutrinos and determine 
the energy spectrum of those sources to learn about the physics that drives their central engines.

In addition, the detection of ultra-high energy (UHE) 
neutrinos ($>100$~PeV) would open a new window into the universe, allowing us to study
distant energetic astrophysical sources that are otherwise inaccessible. 
UHE neutrinos are created as a byproduct of the so-called GZK process~\cite{beresinsky_1969_cosmogenic}, 
the interaction of a cosmic ray (E $>10^{19.5}$~eV) with a cosmic microwave 
background photon~\cite{g,zk}.  
The current best limit on flux of UHE neutrinos comes 
from IceCube below $10^{19.5}$~eV~\cite{icecube2015}, and ANITA above $10^{19.5}$~eV~\cite{anita2}.
Detection of these neutrinos, often called cosmogenic neutrinos, would shed light on the source of the highest 
energy cosmic rays, tell us about the evolution of high energy sources in our universe, and give us information about
the composition of high energy cosmic rays.  We expect neutrinos from the GZK process
to have extremely high energies, above 100~PeV~\cite{ess, kalashev, ABOWY, barger, stanev, kotera}.  
Detection of UHE neutrinos
would also allow us to study weak interactions at center of mass energies $\gtrsim100$~TeV, much 
greater than those available at particle
colliders on Earth, such as the LHC~\cite{hooper_2002_cross_section,Connolly:2011vc,Klein:2013xoa}.
As extremely relativistic particles, UHE neutrinos would enable sensitive tests of 
Lorentz invariance~\cite{gorham_2012_liv} and other beyond Standard Model processes.

There are clear requirements for an ideal high energy neutrino observatory 
driven by the twin science goals of determining the origin and spectrum
of the observed astrophysical 
IceCube signal, and discovering cosmogenic neutrinos to determine the origin of the highest energy cosmic rays 
while also probing particle physics at extremely high energies.
The requirements for a high energy neutrino observatory
are to 1) have the sensitivity in the PeV energy range to measure the observed IceCube 
astrophysical neutrino spectrum and extend the measurement to higher energies, 2) 
have the geometric acceptance at extremely high energies (E $>10^{17}$ eV) 
to detect cosmogenic neutrinos, even in the 
most pessimistic of reasonable neutrino flux models, 3) have the pointing resolution required (sub-degree) 
to determine the origin of the observed neutrinos, and 
4) have the energy resolution (factor $\sim2$) required to 
measure the neutrino energy spectrum at PeV energies and above to learn about the physics 
that drives the sources of high energy cosmic rays and neutrinos.

One promising way to detect the highest energy neutrinos is through the coherent, impulsive radio emission 
from electromagnetic showers induced by neutrino interactions in a dielectric --- the Askaryan effect~\cite{askaryan}.  When a neutrino 
interacts with a dielectric such as ice, an electromagnetic cascade is initiated, and a net negative charge excess
develops in the medium.  This net charge excess moving faster than the speed of light in the medium yields
Cherenkov radiation.  For long wavelength, low frequency emission (frequency $<$1 GHz), the emission is coherent, and for high energy
showers, the radio emission dominates.  
Beam test measurements~\cite{saltzberg_2001_sand,gorham_2005_salt,gorham_2007_ice} confirm that the emitted radio power scales as the square of the particle cascade energy and validate the expected angular emission pattern and frequency dependence from detailed numerical simulations~\cite{zas_1992_zhs}.
A good medium for detection of high energy neutrinos is a large volume
of a dense dielectric with a long radio attenuation length, such as glacial ice, which has an attenuation
length $L_{\alpha} \sim 1$~km~\cite{avva,southpoleice}.

We introduce here a new type of radio detector for high energy neutrinos that will meet these requirements.
In Section \ref{sec:concept}, we review current radio detector techniques and describe the phased radio array concept.
The projected gains in sensitivity using a phased array are presented in Section \ref{sec:results}.
We conclude in Section \ref{sec:conclusion}.

\section{A new experimental approach: an in-ice phased radio array}
\label{sec:concept}
\subsection{Defining a general approach}

A high energy neutrino observatory that achieves the goals outlined in Section~\ref{sec:intro} must have sensitivity over a broad energy range from PeV to $10^{5}$ PeV scales.
To reach the lowest possible energy threshold, the instrument should be located as close to the neutrino
interactions as possible.   
The electric field strength falls as $1/r$ (where $r$ is the distance from the detector to the shower induced
by the neutrino interaction) and also suffers attenuation in the detection medium.
The lowest threshold will be achieved by a set of detectors 
that is directly embedded in a detection medium with a long radio attenuation length,
such as glacial ice.

To achieve both a low energy threshold and increase the effective volume, 
the signals used to trigger the detector should have as high an effective gain as possible.
High-gain antennas produce a higher signal-to-noise ratio (SNR) per antenna compared to low-gain antennas for radio signals aligned with the boresight, 
but have reduced angular coverage and may be impractical to deploy down a narrow borehole.
Combining signals from multiple low-gain antennas in a phased array provides another way to achieve high gain and therefore allow weaker neutrino signals to be detected while also allowing full angular coverage.  
The phased array approach is the topic of this work.

\subsection{Current radio detector techniques}

RICE was a pathfinder experiment for the radio detection of UHE neutrinos and demonstrated the feasibility of many operation-critical technologies~\cite{kravchenko_2012_rice}.
The extremely high energy range, above $10^{19}$~eV, is currently probed by the ANITA high altitude balloon experiment~\cite{instrument}.
The proposed balloon-borne EVA experiment~\cite{eva} would be a novel way to reach the highest
energy neutrinos, above $10^{19}$~eV.
The ARA~\cite{araWhitepaper} and ARIANNA~\cite{arianna} experiments, 
ground-based radio arrays in early stages of development with a small number of stations deployed in Antarctica, 
have energy thresholds $\gtrsim 50$~PeV,
reaching the heart of the cosmogenic neutrino regime. 
This is achieved by placing the detectors in the ice.  

ANITA, ARA, and ARIANNA all use a similar fundamental experimental design: an array of antennas (16 in the case of ARA)
and a data acquisition system comprise a single station.  
The stations are quasi-independent in that each individual station can reconstruct a neutrino event. 
ANITA can be viewed as a single-station experiment in this context.
For ground-based experiments, multiple stations can be positioned several kilometers apart to cover large volumes of 
ice, and the neutrino event rate increases linearly with the number of stations.

In current and previously-deployed experiments, 
a threshold-crossing trigger is used to determine when individual antennas receive an excess in power above 
typical thermal noise~\cite{instrument,ara,arianna_2015}.
If a sufficient number of coincident antenna-level triggers occur within a short time window, a station-level 
trigger is formed, and the antenna waveforms of the candidate neutrino event are digitized and recorded.
Essentially this type of combinatoric threshold-crossing trigger is only sensitive to the amount of power seen by individual antennas as a function of time.
With this triggering approach, the smallest signal that a station can see 
is determined by the gain of each individual antenna (i.e., how much power the antenna sees from signal 
compared to thermal noise).  

\subsection{An in-ice phased array concept}
\label{sec:calc}
We present here a new concept for radio detection of high energy neutrinos that will 
for the first time have sensitivity in the 1~PeV energy range, allowing us to 
characterize the measured IceCube 
astrophysical neutrino spectrum, extend the measurement to higher energies, and achieve 
meaningful overlap with IceCube in energy for energy calibration.  
This new radio detector would also achieve improved sensitivity in the UHE regime per station, 
provide superior pointing resolution at all energies, and provide stronger rejection against 
anthropogenic radio frequency interference compared to currently-implemented radio techniques. 

The key to lowering the energy threshold of a radio experiment and increasing sensitivity at higher energies 
is the ability to distinguish weak neutrino-induced impulsive signals from thermal noise.  
For antennas triggering independently, the amplitude of the thermal noise is solely determined by 
the temperature of the ice and the noise temperature of the amplifiers (a smaller effect than the ice itself).  

Combining signals from many antennas in a phased array configuration averages down the uncorrelated thermal 
noise from each antenna while maintaining the same signal strength for real plane-wave signals (such as
neutrinos).  
If we combine the signals from multiple antennas with the proper time delays to account for the 
distance between antennas, we can effectively increase the gain of the system of antennas for 
incoming plane waves from a given direction. 
Many different sets of 
delays with the same antennas can create multiple effective antenna beam patterns that 
would together cover the same solid angle as each individual antenna but with much higher gain.  
This procedure is called beamforming, and 
is an economical and efficient way to achieve the extremely high effective gain
needed to push the energy threshold down to 1~PeV. 
Such interferometric techniques have been extensively used in radio astronomy~(for a review, see~\cite{thompson}). 

The effective gain $\mathrm{G}_{\mathrm{eff}}$ in dBi or dBd
of the system is determined by the gain of each individual antenna in dBi or dBd, G, and the number of antennas, N, by:
\begin{equation}
\mathrm{G}_{\mathrm{eff}} = 10 \log_{10}(\mathrm{N}\times 10^{\mathrm{G}/10}).
\label{eqn:gain}
\end{equation}

The trigger threshold of radio detectors is typically set by the rate at which antenna waveforms can be digitized and recorded while maintaining a high livetime fraction.
Lower thresholds on the electric field at the antennas correspond to both increased efficiency for neutrino signals and higher trigger rates.
For a phased array, each trigger channel corresponds to a single effective beam, and the station-level trigger is the union of simple threshold-crossing triggers on the individual effective beams of the 
phased array, rather than individual antennas.
In this configuration, the threshold on the electric field could be reduced by roughly the square root of the number of antennas that are phased together while maintaining the same overall trigger rate per trigger channel.
Since the electric field produced at the antenna from Askaryan emission scales linearly with the energy in the particle cascade, this reduction in the effective electric field threshold directly translates into a lower energy threshold for finding neutrinos.
Equivalently, the effective volume of the detector is increased at fixed neutrino energy since events could be detected from farther away.
To minimize the number of trigger channels, thus minimizing complexity and cost, the phased array that provides the
trigger needs to be as closely packed as possible.  Since the overall physical size of the array compared to the
wavelength of radiation determines the angular resolution of the instrument, 
the closer the spacing between antennas, the larger each effective beam
is for a given frequency of radiation,
and the fewer channels are required to cover the same solid angle of ice~\cite{thompson}.

Further sensitivity gains may be possible by recognizing that the antennas that form the trigger 
do not have to be the same antennas used for detailed event analysis.
Indeed, it is advantageous to construct two distinct, co-located
antenna arrays.  The first is the phased array that is as closely packed as possible and provides 
the most sensitive trigger possible (the ``trigger array'').  
The second array is a set of antennas that are spaced as far apart as is reasonably 
possible (many tens of meters) to provide the best pointing resolution and energy resolution possible 
for neutrino events 
(the ``pointing array'').  
Although the station triggers on the compact trigger array, we do not need to digitize and record signals 
from those antennas --
instead, we digitize and record signals from the much farther spaced pointing array based on the timing given
by the trigger.  The compact trigger array also has the benefit that the position of
the antennas only needs to be determined to within a few inches in order to phase the antennas properly.
The distinct sets of antennas that comprise the pointing array and the trigger array 
enable both optimal sensitivity and optimal directional pointing and energy reconstruction
compared to the scenario where one compromises both technical goals 
in order to find middle ground that uses the same antennas for the trigger and pointing.
This two-component station design represents a departure from all previous radio detector neutrino experiments.

The phased trigger array also has the benefit that it can help reject anthropogenic radio frequency interference.  
A low-gain antenna has a very wide beam pattern, and is sensitive to many incoming directions of radiation.  However,
a high-gain antenna, such as is effectively created by the phased trigger array, is highly directional.  Since
this configuration would have many effective high-gain beams, each pointed in a different direction, at any given time 
we can easily mask out directions where there is man-made interference from the trigger.  
Since man-made noise tends to come from specific incident directions corresponding to the location
of sources of noise, the ability to mask out directions from the trigger improves background rejection and
allows the threshold of the trigger to be set by thermal noise for a larger fraction of time, rather than
the threshold increasing when more man-made radio frequency interference is present.

Equation~\ref{eqn:gain} holds for the case where the signal is amplified with a low-noise amplifier for
each antenna before the beams are formed (i.e., the noise contribution from each channel is the same and no significant
noise is introduced after the beams are formed). 
If the loss in the system is low enough, 
the first stage of amplification could be performed after the beams have been formed, so
each beam has a noise contribution from only one amplifier.  This improvement would come
at the cost of including any noise from loss in the cables and the beamforming components.  In a typical case,
the array views 250~K ice and the front-end system has a temperature of 70~K.  Even if the beamforming hardware
were lossless, the improvement in overall system temperature is small, approaching 10\% improvement only after 
phasing 400~antennas, since the noise from the ice dominates the noise from the amplifier system.  Despite 
only modest potential improvements, this technique may ultimately allow for fewer front-end amplifiers
to be used, which would reduce the cost of the array.  

An interferometric technique has been previously developed and used in post-processing analysis for 
fast, impulsive radio signals from neutrino interactions.  
This interferometric technique was first developed for and applied 
to the ANITA experiment~\cite{interferometry}, and has since been applied
to other experiments~\cite{ara,ara2015}.  Interferometric techniques have also been used to search for extremely high 
energy neutrinos, above $10^{22}$~eV, using the Westerbork Radio Synthesis Telescope and
the moon as a target~\cite{lofar}.

There have been efforts directed toward reconfiguring the triggering scheme of currently-deployed
or soon-to-be deployed experiments, such as ANITA and ARA, to instead use real-time correlation triggering
after 3-bit digitization of the Nyquist-sampled waveforms~\cite{ritc}.  This technique is under
development, and is expected to be used in the upcoming fourth flight of ANITA.
There are two reasons why one should consider doing the beamforming in hardware. The first is that it leaves open 
the possibility of doing the signal amplification after the beams are formed to reduce the noise
contribution from the front-end electronics.  The second is that information is lost in the 3-bit digitization that 
is preserved by doing the full phasing in hardware.  

As described above, the optimization of the geometry of a station 
should be different for a phased radio array compared to currently-deployed instruments to fully exploit
the power of the beamforming technique.  We explore here two scenarios: the first is a 16-channel station
much like the deployed stations for the ARA experiment but with a different geometry and with the phased array trigger
implemented, and the second is a 400-channel station
that would achieve a 1~PeV threshold.

\subsection{Example: a 16-channel station}

\begin{figure}
  \centering
  \includegraphics[width=12cm]{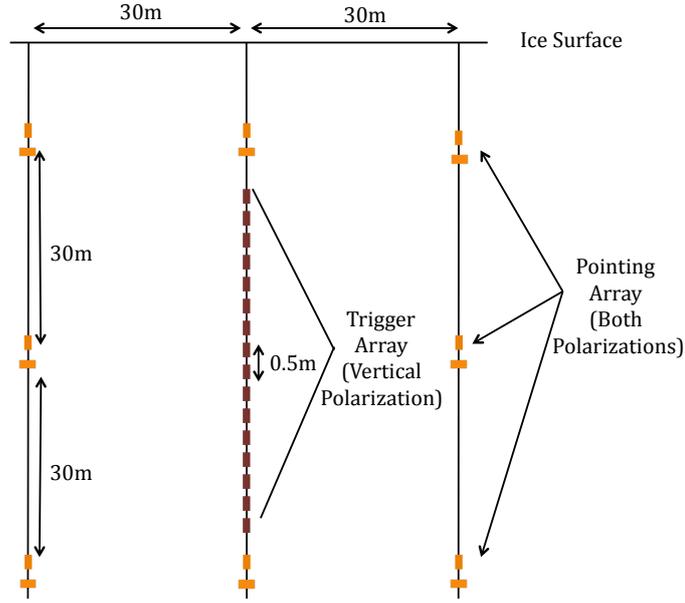}
  \caption{An example station layout for a 16-antenna phased trigger array and accompanying pointing array.}
  \label{fig:station}
\end{figure}
\begin{figure}
  \centering
  \includegraphics[width=12cm]{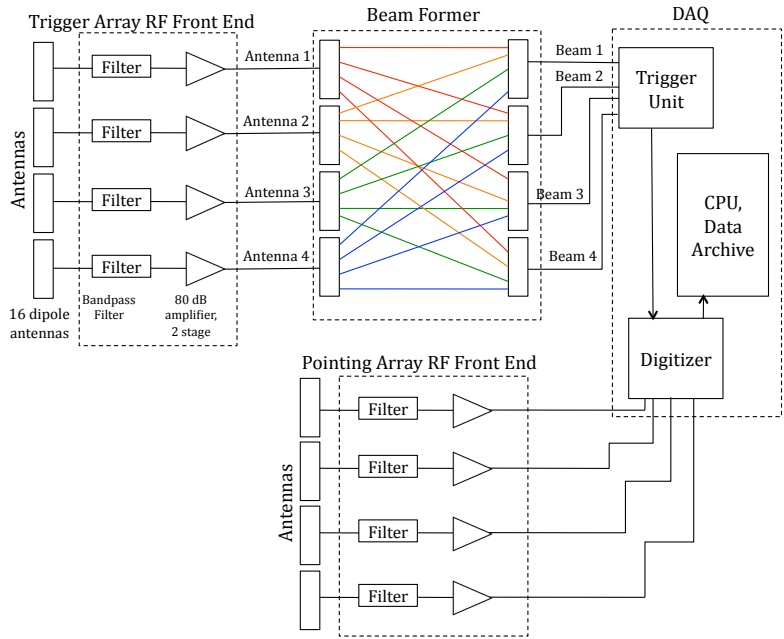}
  \caption{An example layout of the radio frequency chain of a 16-antenna phased trigger array and 
    accompanying pointing array. For simplicity, not all channel paths are depicted.}
  \label{fig:electronics}
\end{figure}
\begin{figure}
  \centering
  \includegraphics[width=11cm]{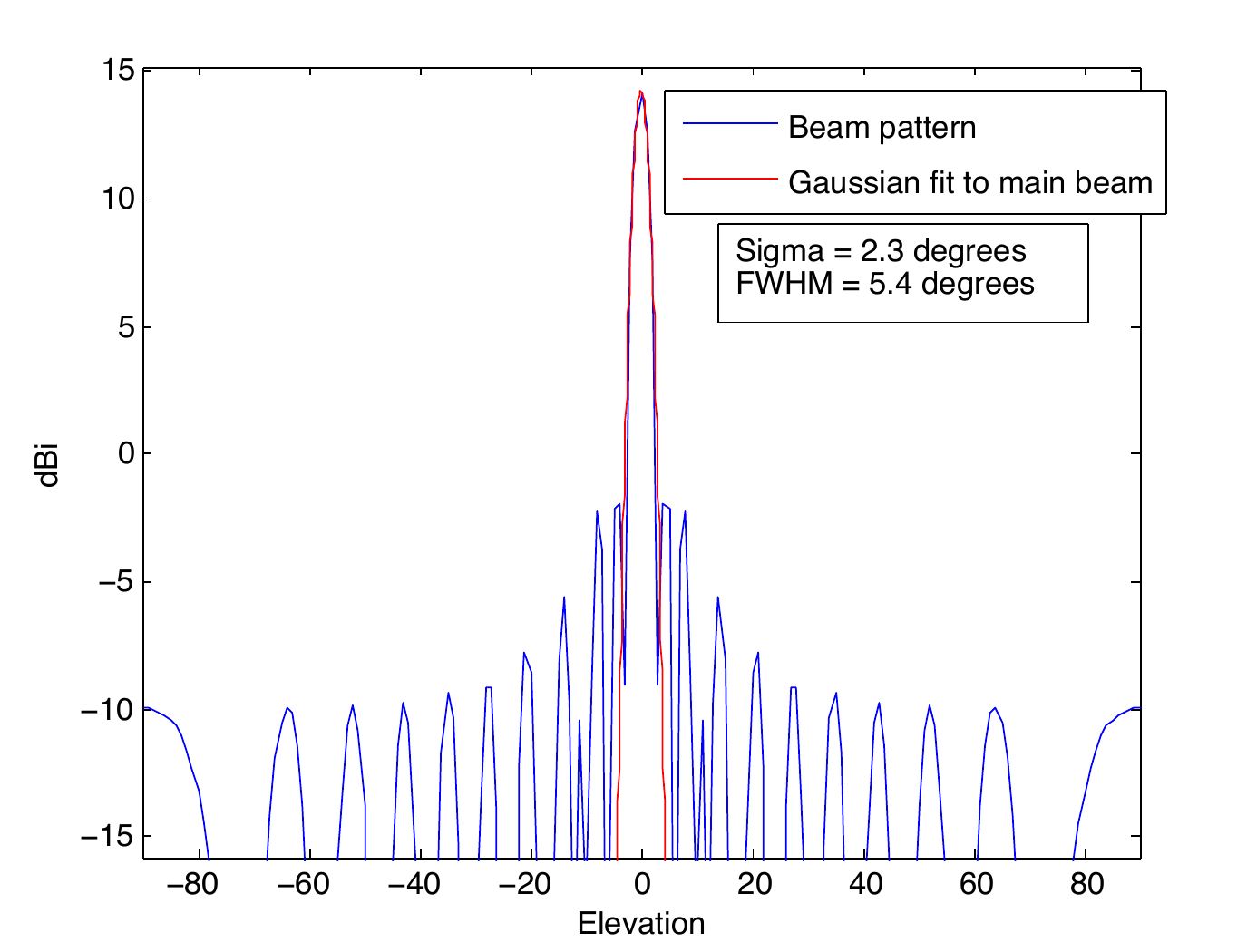}
\caption{The beam pattern for one trigger channel at 200 MHz for the configuration 
  shown in Figure~\ref{fig:station}.}
\label{fig:pattern}
\end{figure}

We introduce here the example of a 16-channel station that uses dipole-like antennas and is 
deployed down boreholes beneath the firn layer of a deep glacier.  
Possible sites for such a station, or set of stations,
include South Pole, where the array would
be $\sim200$~m below the surface and the ice is $\sim2.8$~km deep~\cite{koci,price}, and Summit Station
in Greenland, where the array would be $\sim100$~m below the surface and the ice is $\sim3.0$~km deep~\cite{gow_1997_gisp2}. 

We simulate a configuration where the antennas are deployed 
down boreholes below the firn.  Our simulations confirm the finding from
ARA~\cite{araWhitepaper,ara} that due to the changing index of refraction in the firn as the
glacial ice transitions to snow, incident radio emission is often deflected downwards away from a detector on the 
surface and is therefore undetected (see~\ref{sec:ray-tracing} for details).
In contrast, receivers deployed below the firn layer are less affected by refraction and the increase in effective
volume compared to a surface configuration is large (factors of 3 to 10 depending on the firn profile).
The benefits of borehole-deployed antennas make them a cost effective approach.

One possible station layout is shown in Figure~\ref{fig:station}.  
A trigger array is constructed of 16 dipole antennas strung vertically
down one borehole as close together as possible.  This configuration would naturally be sensitive to vertically-polarized signals,
although one could combine signals from orthogonally-polarized antennas to create an unpolarized trigger.
The pointing array would be constructed of additional antennas, with 
both horizontal and vertical polarization sensitivity, and would require at least two additional boreholes to uniquely determine
the incident direction, timing, and polarization of the radio emission 
and thus the incoming direction of the neutrino.  We would only need to digitize the signals from the pointing 
array antennas.  An example of the radio frequency chain for a 16-antenna station is shown in Figure~\ref{fig:electronics}.

The effective gain of such a 16-channel trigger array of dipole antennas calculated using Equation~\ref{eqn:gain} 
is 14.2~dBi (compared to 2.15~dBi
for a single dipole), which corresponds to a factor of $\sim$4 in electric field threshold. 
The effective beam at 200~MHz 
for one trigger channel is shown in Figure~\ref{fig:pattern}
as a function of elevation. The FWHM of the beam 
of a single trigger channel is $5.4^\circ$ in elevation with complete azimuthal coverage for this configuration.
By adjusting the delays among antennas in different trigger channels, we can cover the relevant range of solid 
angle for incoming emission from visible neutrino events with only $\sim$15 beams.
Antennas with a moderately higher gain
that still cover the relevant range of solid angles would lead to a higher effective gain
of the system.

\subsection{Achieving An Energy Threshold of 1 PeV}

The phased array design is scalable and can be configured to achieve an even lower energy threshold.
For example, a phased array with 400 dipole antennas would have an effective gain of 28.2~dBi, 
and would push the electric field threshold down by a factor of $\sim$20
compared to currently-implemented techniques.  For large numbers of antennas, the single-borehole configuration
for the trigger array is no longer optimal.  To keep the size of the trigger array compact, trigger antennas
would be deployed down multiple boreholes.
Phased arrays with thousands of channels are possible.

Since the spectrum and angular distribution of Askaryan emission is largely independent 
of the neutrino energy~\cite{alvarez-muniz_2012_hadronic}, the same experimental design principles and simulation methods
are valid over the large range of neutrino energies we consider.

\section{Comparison with current experimental techniques using an independent Monte Carlo simulation}
\label{sec:results}

  \begin{figure}[h]
      \begin{center}
        \includegraphics[width=12cm]{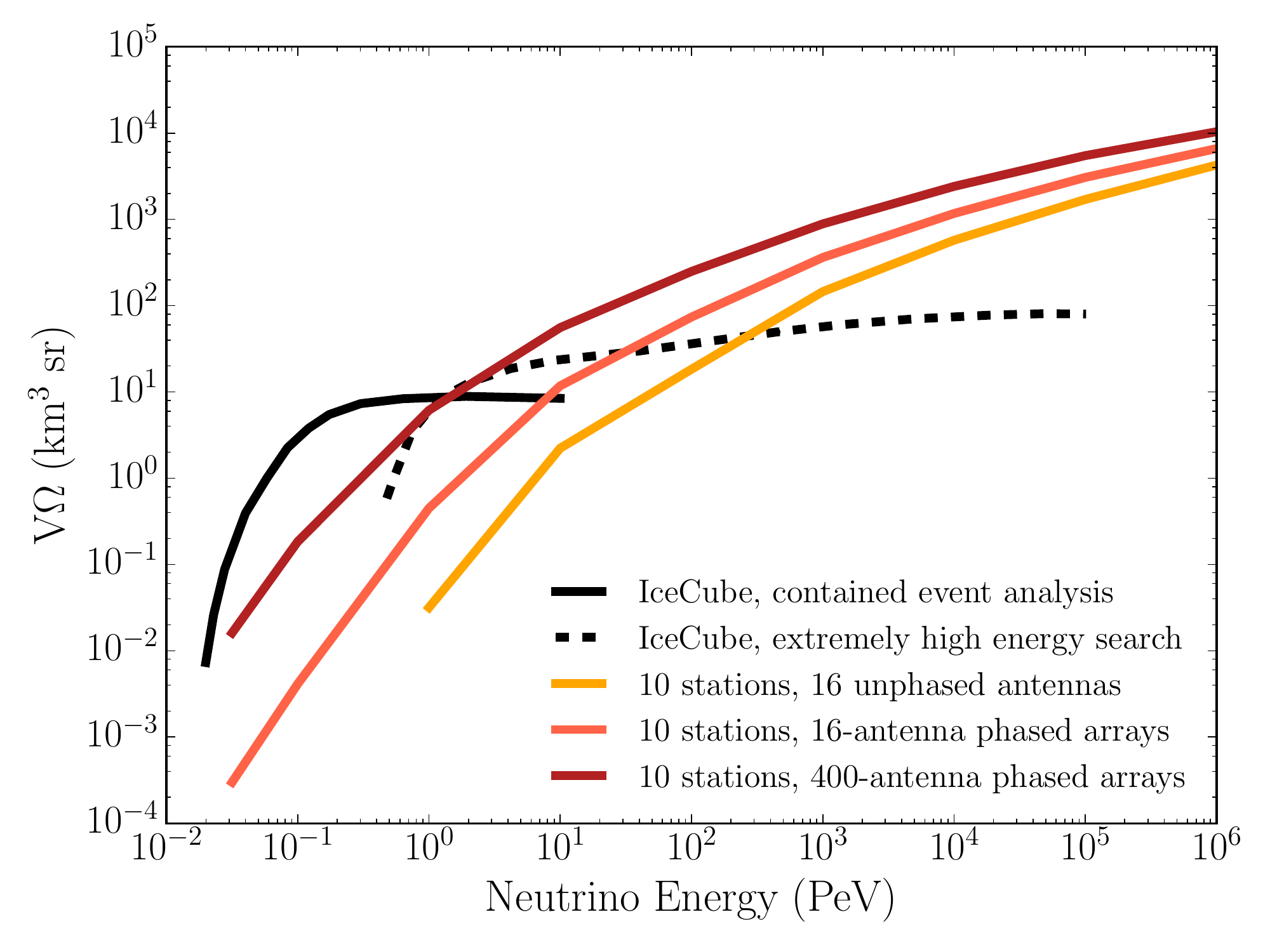}
    \end{center}
    \caption{
      Effective Volume vs. Energy for 10 stations installed 100~m below the surface at Summit Station,
      Greenland. The yellow line is for 16-channel stations with no phasing, 
      the orange line is for similar
      stations but with phasing, and the red line is for 400-antenna phased array stations.
      For each radio array configuration, the volumetric acceptance is presented at the trigger level.
    Black curves indicate the volumetric acceptance for two difference analyses with IceCube optimized for different energy ranges \cite{icecubeContained,icecubeEHE}.} 
    \label{fig:simulation}
  \end{figure}

\begin{figure*}
\centering
\includegraphics[width=7cm]{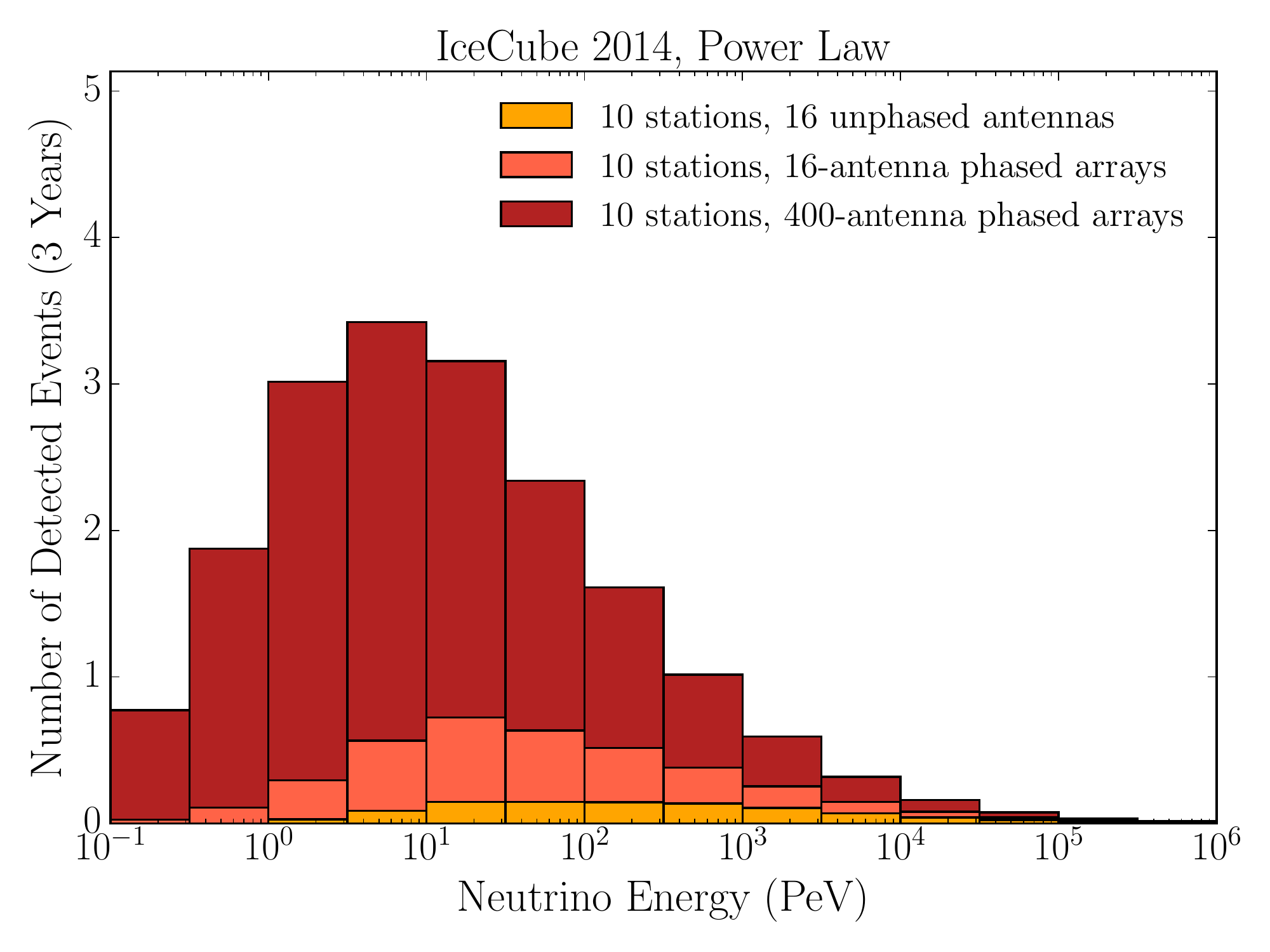}
\includegraphics[width=7cm]{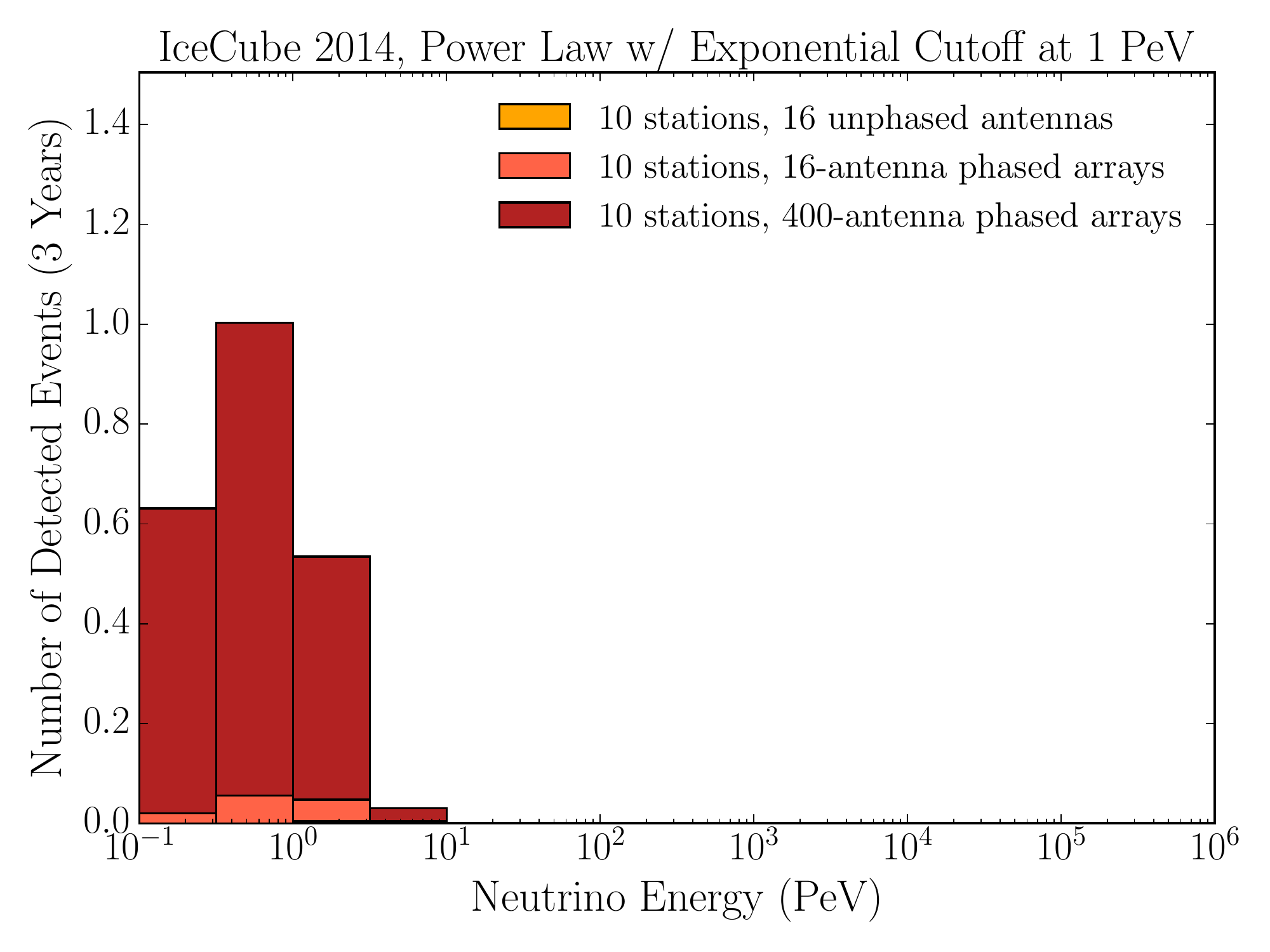}
\includegraphics[width=7cm]{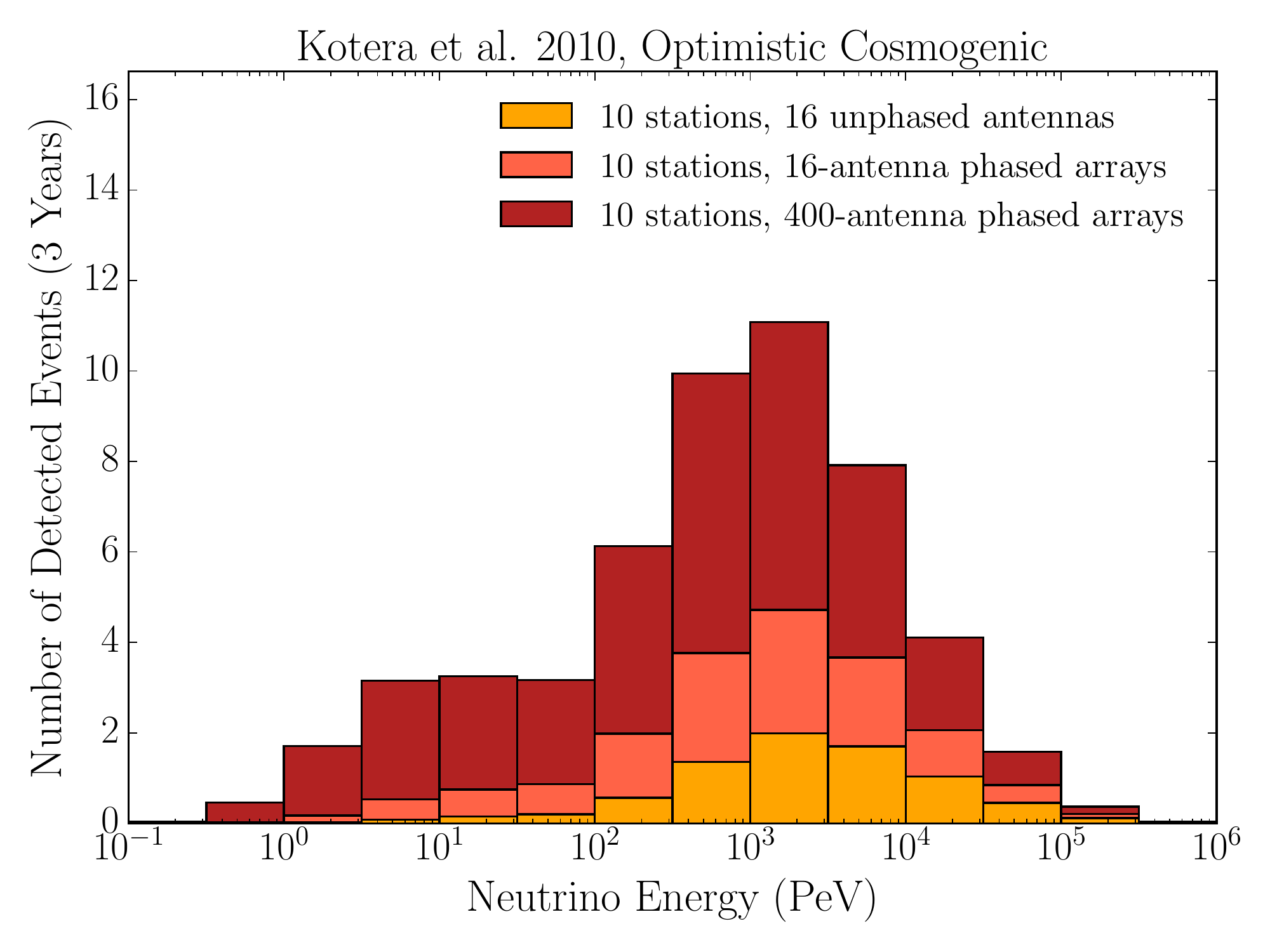}
\includegraphics[width=7cm]{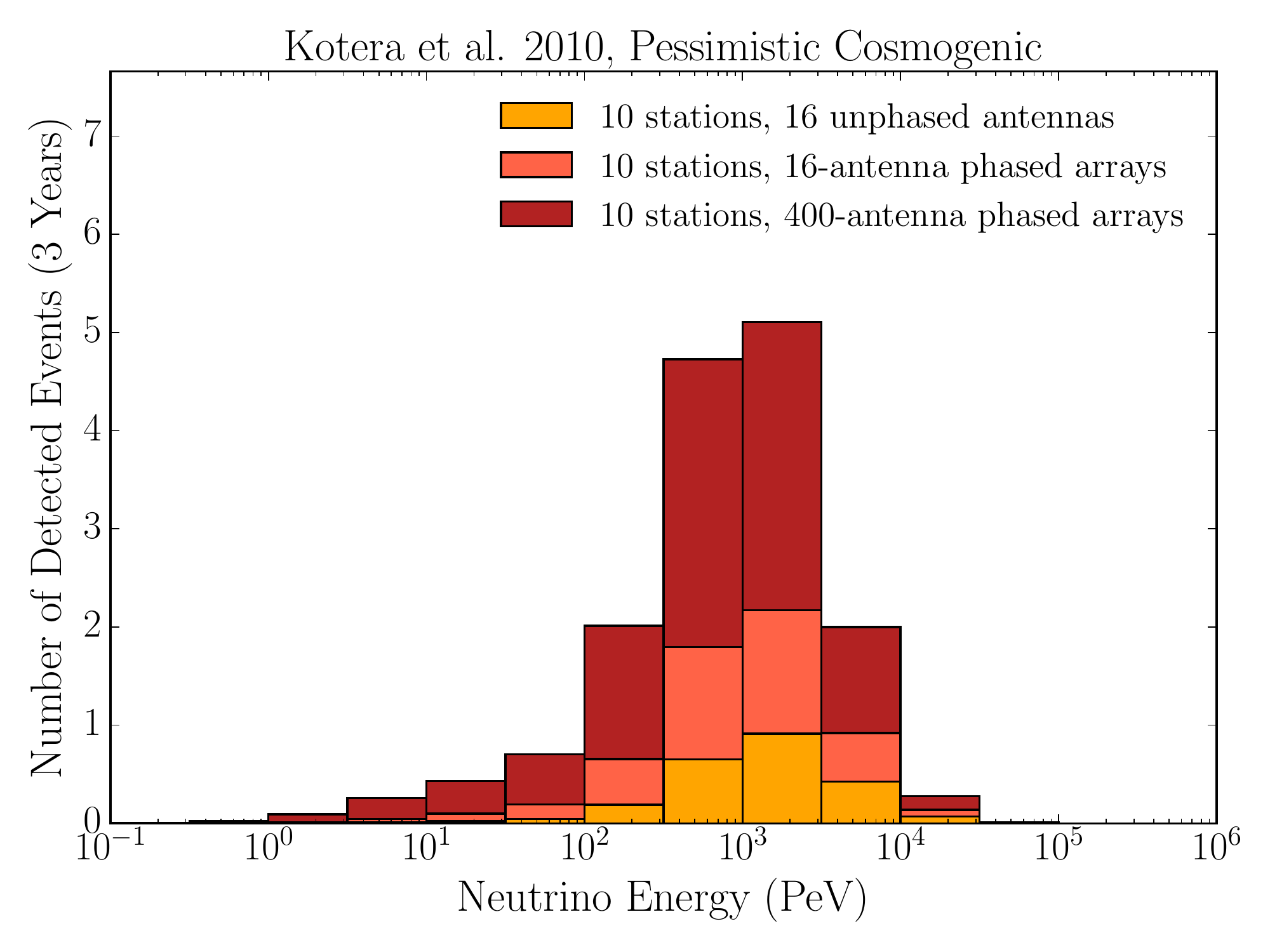}
\caption{Triggered Event Rates vs. Energy for a variety of neutrino models.  Triggered event rates for three years of observation 
  for 10 stations installed 100~m below
  the surface at Summit Station, Greenland. 16-channel stations with no phasing 
  are shown in yellow, 16-channel stations with phasing are shown in orange, and stations with 400 phased
  antennas are shown in red.
  The top two panels show event rates based on 
  two possible neutrino spectra based
  on the IceCube observed neutrino flux~\cite{bigBird}.  
  The top left panel is an $\mathrm{E}^{-2.3}$ power law, and the top right panel
  is an $\mathrm{E}^{-2.3}$ power law with an exponential cutoff at 1~PeV.  
  The bottom two panels show event rates based on 
  optimistic and pessimistic cosmogenic fluxes~\cite{kotera}.}
\label{fig:events}
\end{figure*}
  \begin{figure}[h]
      \begin{center}
        \includegraphics[width=12cm]{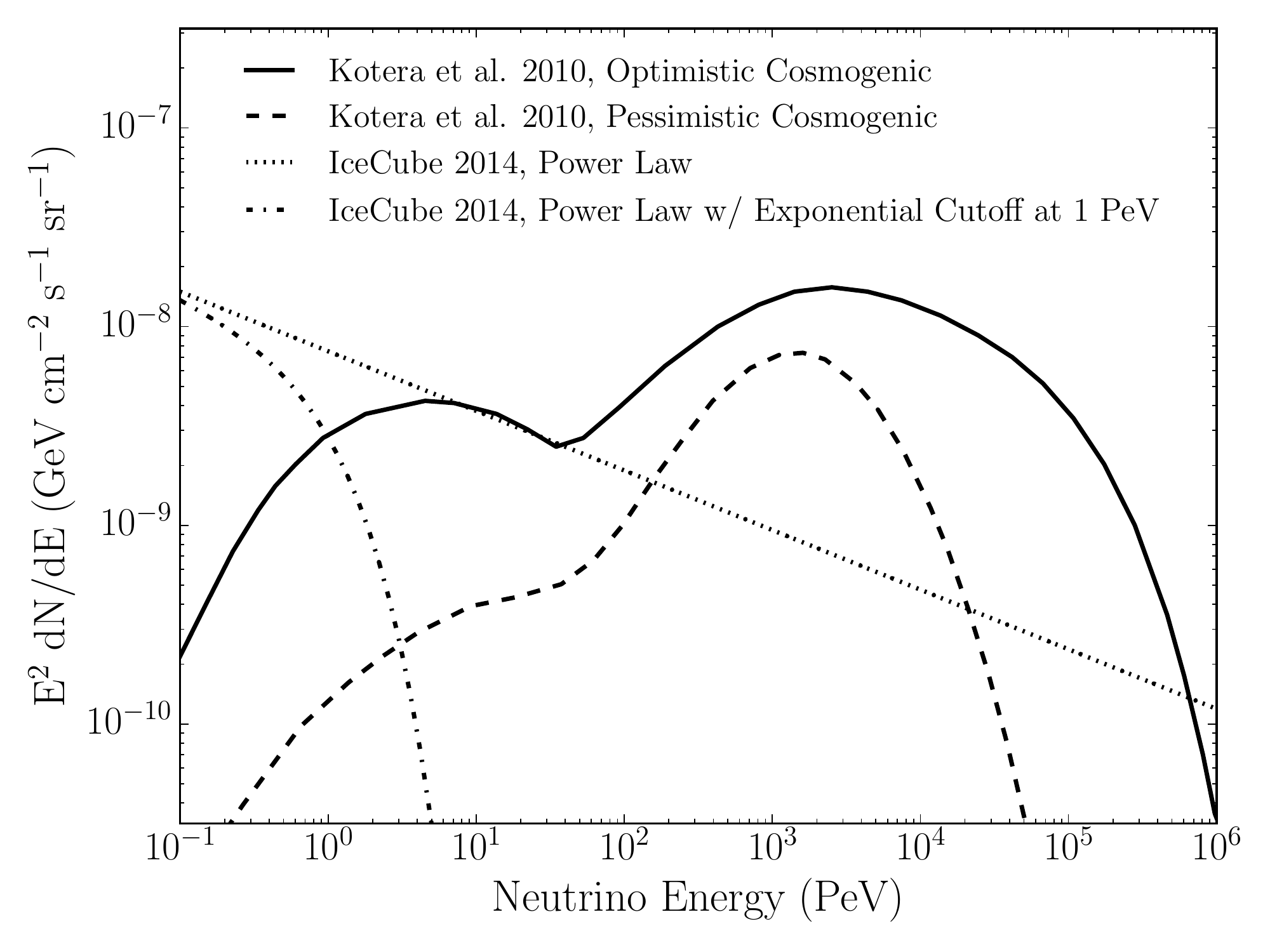}
    \end{center}
    \caption{
      Flux models used for the predictions shown in Figure~\ref{fig:events}.  
      Shown are two possible neutrino spectra based on the IceCube observed neutrino 
    flux~\cite{bigBird}: an $\mathrm{E}^{-2.3}$ power law and an $\mathrm{E}^{-2.3}$ power law with an 
    exponential cutoff at 1~PeV.  
    Also shown are optimistic and pessimistic cosmogenic fluxes~\cite{kotera}.} 
    \label{fig:models}
  \end{figure}

We have developed a Monte Carlo simulation package to quantify the acceptance of various radio detector configurations, 
and more specifically, investigate the advantages of a phased array design.
The simulation formalism and assumed physics input are described in~\ref{sec:formalism}, 
and validation studies are presented in~\ref{sec:validation}. 
Rather than focusing on specific antenna designs, signal processing chains, and analysis algorithms, we have kept the simulations general by defining signal detection thresholds in terms of the electric field strength arriving at the antennas.
In this case, different station configurations correspond to different electric field thresholds, as described above. 
The parameters chosen to define the electric field threshold of our baseline station configuration are discussed in~\ref{sec:threshold}.
Depending on the particular system deployed,
the overall results could shift up or down (by less than a factor of two), 
but the relative comparisons between configurations are valid.

Figure~\ref{fig:simulation}
shows the volumetric acceptance of a 10 station detector with antennas 100~m below the surface at Summit Station
as a function of energy for three different configurations: 16-channel stations with no
phasing (yellow), 16-channel stations with phasing (orange), and 400-antenna phased array stations (red).  
See~\ref{sec:formalism} for a description of how effective volume is calculated.
We also show for comparison 
the acceptance corresponding to two IceCube analyses optimized for different energy ranges, calculated using the 
effective area given in~\cite{icecubeContained,icecubeEHE} 
and the neutrino interaction cross section given in~\ref{sec:formalism}.
The IceCube curves in the figure include analysis efficiency, whereas the results of our simulation do not, so the
curves are not directly comparable.
The standard method, shown with the yellow line in
Figure~\ref{fig:simulation}, is comparable to the approach of the ARA experiment but with only 10 stations~\cite{araWhitepaper}.

For this study, 
we have chosen to simulate the experiment at Summit Station in Greenland.  Moving the detectors to South Pole at a depth
of 200~m below the surface would
increase the acceptance by $<20$\% due to a longer radio attenuation length in ice at the South Pole~\cite{avva}.
We have chosen to simulate 10~stations for this study, but we note that the acceptance changes linearly with the 
number of stations, since stations trigger independently.  

Figure~\ref{fig:events} shows the number of events detected as a function of energy for the same three detector
configurations shown in Figure~\ref{fig:simulation} for a variety of astrophysical and cosmogenic 
models.  We show event rates for an $\mathrm{E}^{-2.3}$ power law
based on IceCube observations~\cite{bigBird}, 
an $\mathrm{E}^{-2.3}$ power law with a 1~PeV exponential cutoff~\cite{bigBird}, and optimistic and pessimistic cosmogenic
models~\cite{kotera}.  The flux models used for Figure~\ref{fig:events} are shown in Figure~\ref{fig:models}.
Table~\ref{tab:events} summarizes the expectation values for the total number of events detected with each detector configuration for each model.  A harder spectrum for PeV-scale neutrinos would yield a higher event rate.

For the most pessimistic cosmogenic neutrino flux models (not shown in Figures~\ref{fig:events},~\ref{fig:models}, or Table~\ref{tab:events}) with no source evolution and a pure iron composition for UHE cosmic rays~\cite{kotera}, 
the expected event rate with 10 stations of 400 phased antennas each is $\sim0.1$ event per year.
However, recent measurements with Pierre Auger Observatory and the Telescope Array 
disfavor a significant iron fraction in the cosmic ray composition at energies up to 
and even exceeding $10^{19.5}$~eV~\cite{auger_2014_composition,telescopeArray}, 
so we do not discuss these cosmogenic models further in this work.

By phasing the antennas in a 16-antenna array, we have improved the UHE neutrino 
event rate by more than a factor of two over the non-phased case, 
and extended the sensitivity to lower energies.
The 16-antenna phased configuration achieves a low enough energy threshold to distinguish 
an $\mathrm{E}^{-2.3}$ power law extrapolation of the observed IceCube spectrum from one that has a
cutoff at the PeV scale.  Phasing more antennas lowers the threshold even further, and makes marked 
improvements in event rates, especially at low energies (see the 400-antenna configuration in Table~\ref{tab:events} 
and Figure~\ref{fig:events}).  Increasing the number of stations results in 
a linear increase in the expected number of events.

As is evident in Figure~\ref{fig:simulation}, 10 stations of 16-antenna phased arrays have a larger acceptance 
than IceCube above 30~PeV, and the acceptance grows faster with energy 
than IceCube, so that by $10^{18}$~eV the acceptance 
is an order of magnitude more than IceCube.  10 stations of 
the 400-antenna phased arrays have a larger acceptance 
than IceCube above 1~PeV.

\begin{table*}
\begin{center}

\begin{tabular}[c]{|l|c|c|c|c|}
\hline 
Station Configuration & Power Law & Power Law & Optimistic & Pessimistic\\
& & with Cutoff & Cosmogenic & Cosmogenic \\
\hline
16-antenna & 0.9 & 0.0 & 7.7 & 2.3 \\
16-antenna, phased & 3.8 & 0.1  & 19.6 & 6.0 \\
400-antenna, phased & 18.4 & 2.2 & 52.9  & 15.6 \\

\hline
\end{tabular}
\end{center}
\caption[]{\label{tab:events}Expectation values for the total number of triggered events in 3 years for 10 stations in different configurations
  for spectra based on IceCube observations~\cite{bigBird} and for cosmogenic models~\cite{kotera}.
 }
\end{table*} 

\section{Conclusions}
\label{sec:conclusion}
We have described a new concept for an in-ice phased radio array that is designed to achieve sensitivity to
the astrophysical neutrino flux at 1~PeV and above, provide energy overlap with IceCube for calibration,
and discover cosmogenic neutrinos in an efficient way.
It is worth noting the scalability of the radio technique for increasing acceptance at all energies.
The acceptance increases linearly with the number of stations, and further gains can be realized by phasing more antennas in each station, particularly at PeV neutrino energies.
An array of 100 stations each with 400-antenna phased arrays could detect hundreds of 
neutrinos at PeV energies and above each year.  
Such an experiment would revolutionize our view of the high-energy universe.

\acknowledgments

This work was supported by the Kavli Institute for Cosmological Physics at the University of Chicago.
Computing resources were provided by the University of Chicago Research Computing Center.  Part of this research
was carried out at the Jet Propulsion Laboratory, California Institute of Technology, under a contract with the 
National Aeronautics and Space Administration.
We would like to thank A. Connolly, P. Gorham, D. Saltzberg, and S. Wissel for useful conversations and guidance.

\appendix
\section{Simulation formalism}
\label{sec:formalism}

We used numerical Monte Carlo simulations to evaluate the acceptance of different array configurations considered in this work.
The individual stations of the array were assumed to be widely spaced such that a given neutrino interaction is unlikely to be detected with multiple stations.
In this limit, the stations act independently and it is possible to simulate the response of a single station and multiply the derived acceptance by the total number of stations in the array to obtain the total acceptance (see, e.g.,~\cite{araWhitepaper}).
We also assumed that the ice properties depend only on depth such that a cylindrical coordinate system with azimuthal symmetry can be defined.
To provide a concrete example, we used Summit Station, Greenland as the site of the envisioned radio array and considered stations at a depth of 100~m, i.e., just below the firn layer \cite{gow_1997_gisp2}.

\subsection{Ray-tracing}
\label{sec:ray-tracing}

\begin{figure}[h]
  \begin{center}
    \includegraphics[width=7cm]{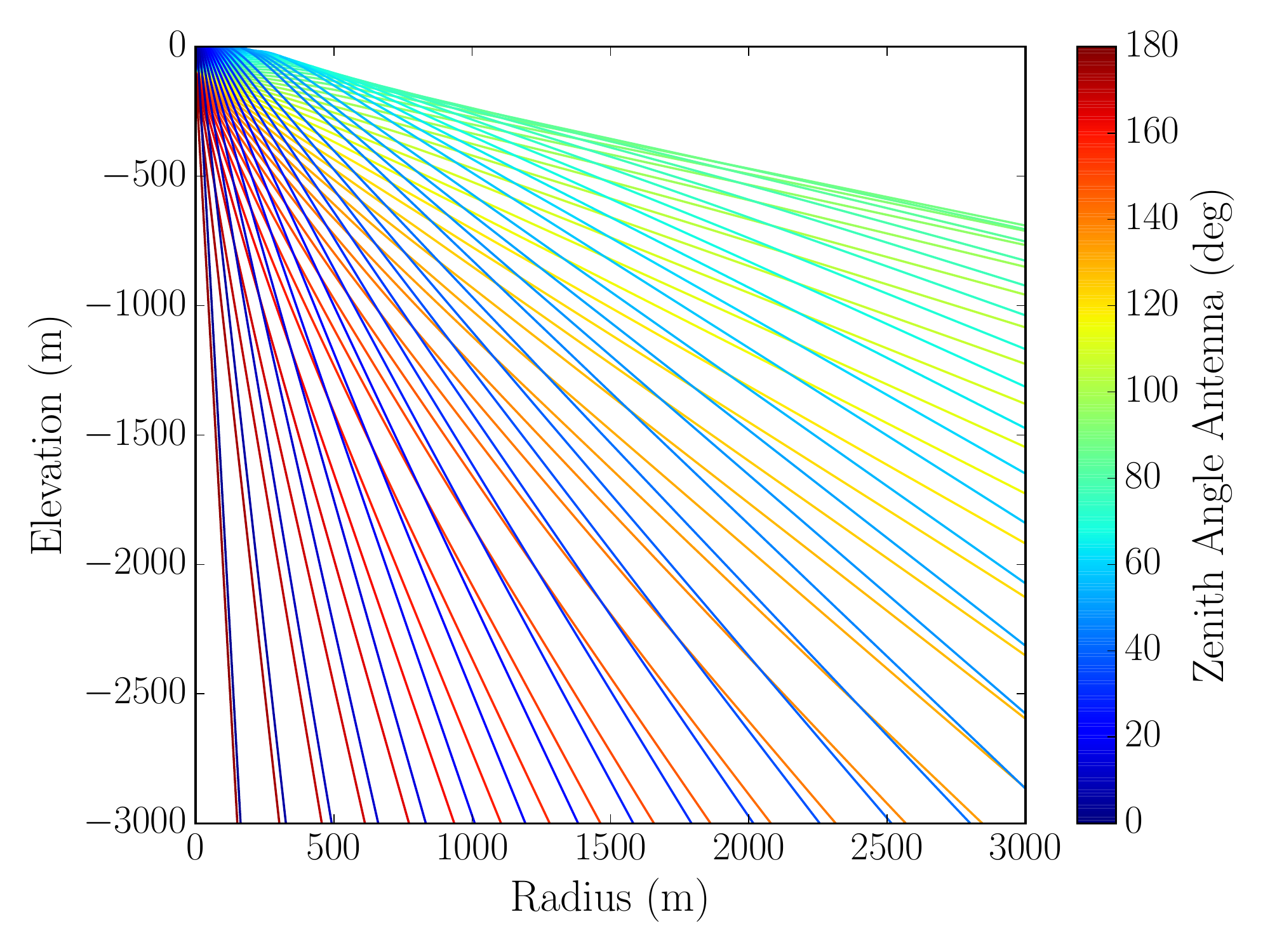}
    \includegraphics[width=7cm]{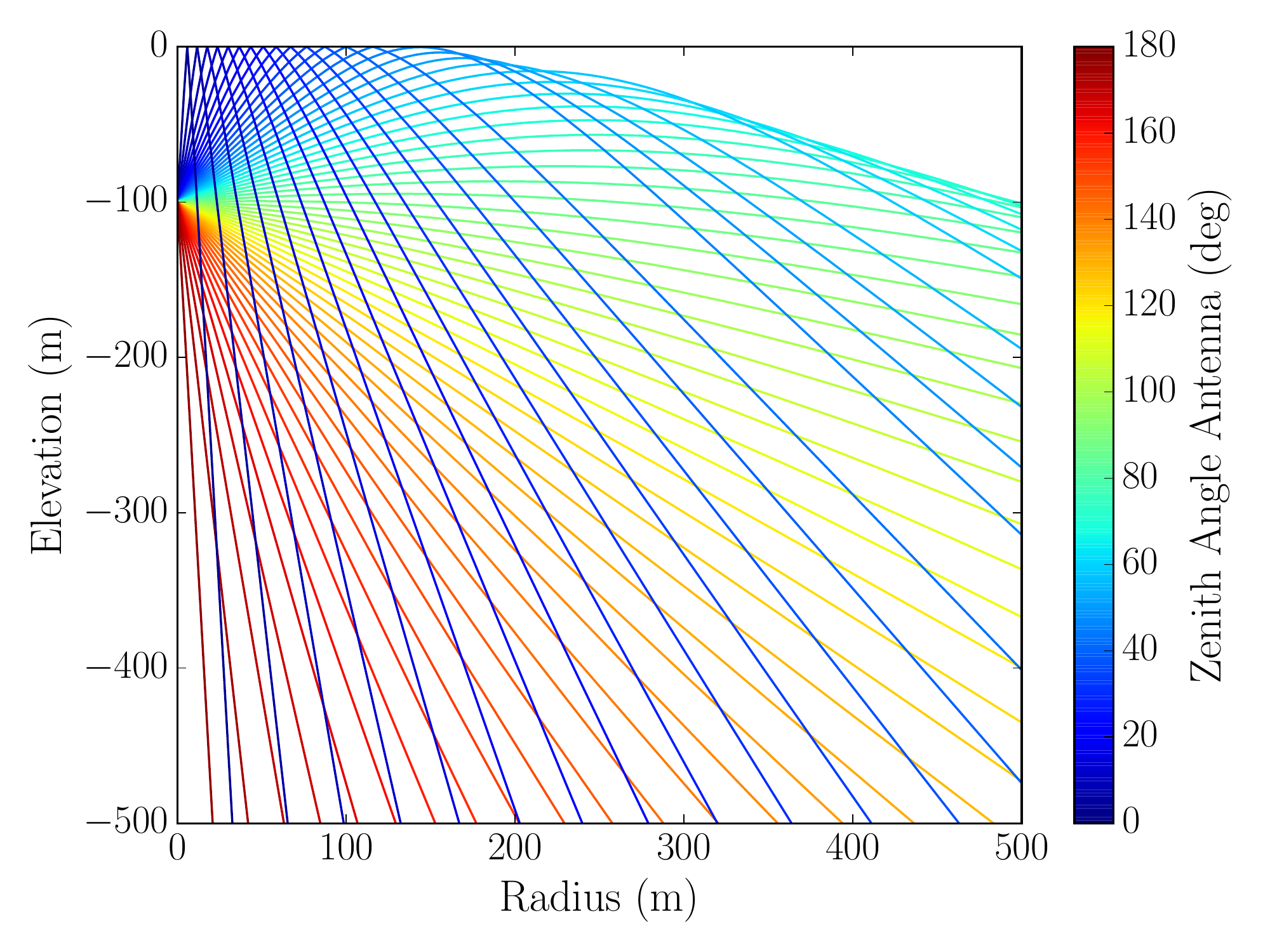}
  \end{center}
  \caption{A ray-tracing diagram for antennas 100~m below the surface of the ice at Summit Station in Greenland.
    The left shows the full depth of the glacier (3000~m), and right is a detailed view close to the station.
    For a given zenith angle of the ray at the antenna, there are three solutions for the true 
    incident radiation direction as discussed in the text.  
    Including waves that reflect off of the ice-rock interface at the bottom of the glacier doubles the number of solutions. 
    The second set of solutions are not shown here.    
    The color indicates the zenith angle at which the antenna sees the ray.} 
  \label{fig:ray}
\end{figure}

The simulations were performed in two stages for computational efficiency.
First, we generated a library of ray-tracing solutions for radio waves propagating through the ice sheet.
Rays originating at the particle cascade must intersect the station in order to be detected.
This boundary condition implies that a complete set of viable ray-tracing solutions can be found by following rays launched at varying zenith angles from the station in the time-reverse sense.
The trajectories are curved due to the varying index of refraction in different layers of the ice, and reflections can occur both at the ice-air interface at the top of the glacier and the ice-rock interface at the base of the glacier.
Measurements of the firn density profile (upper 100~m) at Summit Station show that the index of refraction at $\sim300$~MHz is correlated to the vertical density profile determined from ice cores through the relation $n = 1 + k \rho$ with $k = 8.45 \times 10^{-4}$~kg$^{-1}$~m$^{3}$ \cite{arthern_2013_firn}.
We applied the same formula to density profiles from GISP2 ice core data \cite{gow_1997_gisp2} to estimate the index of refraction throughout the ice sheet.
Figure \ref{fig:ray} shows ray-tracing solutions for a station located at a depth of 100~m. 
The density profile changes rapidly in the firn layer of the ice sheet and therefore ray bending is expected to be most pronounced in the upper $\sim100$~m of the ice sheet.
This effect results in a ``firn shadow'' region that is inaccessible except through reflections at the bottom of the ice sheet~\cite{allison_2009_iceray}.

At short propagation distances from the station, there are three distinct sets of continuous solutions that define regions in which different classes of ray-tracing solutions exist.
The first class involves reflections off the ice-air interface.
The latter two classes involve rays that do not reach the surface, but which can cross over each other due to different amounts refraction in the firn.
It is useful to distinguish between these two classes to recognize regions where two viable ray-tracing solutions corresponding to different zenith angles exist.
The three solution classes are doubled when also considering rays that reflect off the bottom of the ice sheet, resulting in a total of six solution classes.

At each step along a given ray-tracing solution, which are spaced from 1~ns to 5~ns in time depending on the differential index of refraction, we record the location, propagation distance, direction of propagation, and the cumulative attenuation to the station separately in horizontal and vertical polarization.
We used the radio attenuation length measurements at Summit Station from~\cite{avva}.
Fresnel conditions are applied to compute the reflected power at the ice-air and ice-rock interfaces.

\subsection{Sampling neutrino interactions and electric field production}

In the second stage of the simulations, we randomly generated neutrino interaction geometries within a cylindrical target volume centered on the station and used the ray-tracing library described above to compute the electric field that would be produced at the antennas.
Neutrino interaction vertices were selected uniformly throughout the target volume and arrival directions were uniformly distributed over the celestial sphere ($4\pi$ sr).
For computational efficiency, the size of the target volume, $V_{\rm sim}$, was enlarged for the highest neutrino energies and contracted for lower neutrino energies.
In all cases, the depth was matched to the thickness of the ice sheet (3000~m) and the radius was chosen to be large enough to encompass all detectable events.

We focused on neutrino-induced hadronic cascades when computing the Askaryan emission.
Muons and tau leptons produced in charged-current interactions are expected to deposit their energy in elongated tracks that would be difficult to detect~\cite{lai_2014_flavor}, and the electromagnetic cascade from charged-current interactions of UHE electron neutrinos can be elongated due to the Landau-Pomeranchuk-Migdal (LPM) effect~\cite{landau_1953_lpm,migdal_1956_lpm,zas_1992_zhs}, resulting in a narrow Cherenkov emission cone~\cite{kravchenko_2006_rice}.
Hadronic cascades are less affected by elongation due to the LPM effect, and the width of the Cherenkov cone is therefore less dependent on neutrino energy~\cite{alvarez-muniz_1998_hadronic,alvarez-muniz_2012_hadronic}.
In addition, simulations of the LPM effect at higher energies have been shown to produce cascades with interfering pulse trains rather than elongated pulses~\cite{alvarez-muniz_2011_askaryan}. 
With the high gain provided by an in-ice phased array, these signals could be more readily differentiated from the thermal noise background.
Estimating the increase in acceptance due to these signals will be the subject of a future study.
Either way, the approach taken here is conservative since the LPM effect will only become dominant for electromagnetic showers with energies $>10^{18}$~eV.

For each candidate neutrino event, we determined viable ray-tracing solutions by interpolating between rays within each respective class of solutions in the ray-tracing library (see above).
Next, for each viable ray-tracing solution, we used the analytic parametrization of~\cite{lehtinen_2004_forte} to compute the Askaryan emission at the unique angle between the neutrino propagation direction (parallel to the cascade development) and the ray connecting the cascade to the station.
We used a characteristic length scale for the shower of 1.5~m, as suggested by~\cite{lehtinen_2004_forte}.
When computing the fraction of the incident neutrino energy that is visible in the cascade, we used the analytic parametrizations for the inelasticity (Bjorken-$y$ parameter) from~\cite{Connolly:2011vc}, which corresponds to the energy in the hadronic cascade.
The inelasticity distributions for neutral current and charged current interactions of both neutrinos and anti-neutrinos are expected to be nearly indistinguishable in the UHE range.
For simplicity, we drew values from the inelasticity distribution for charged-current neutrino interactions.

In the UHE regime, the neutrino-nucleon interaction cross section is sufficiently large that a significant fraction of upward-going neutrinos interact while passing through the solid body of the Earth and are therefore not detectable.
We used the analytic parametrizations for the interaction cross section from~\cite{Connolly:2011vc} and the Preliminary Reference Earth Model (PREM) profile for the Earth's density~\cite{prem} to compute the probability $p_{\rm Earth}$ that neutrinos survive passage through the Earth as a function of their energy and incoming zenith angle.
The Earth passage probability takes into account the distance traversed through the ice sheet on the way to interaction vertex. 
For the purpose of computing the neutrino interaction probability, we used the total interaction cross section for neutrinos (as opposed to anti-neutrinos).
In the energy range considered here, the total cross section for neutrinos is a factor of 1.06 to 1.25 larger than for anti-neutrinos.
The current simulations do not include the cross section enhancement from the Glashow resonance, $\bar{\nu_e} e \rightarrow W^{-}$, which would lead to a further enhancement in acceptance for electron anti-neutrinos of energy $\sim6.3$~PeV \cite{glashow_1960_resonance,berezinksi_1977_resonance,berezinsky_1981_resonance}.

The process described above effectively samples the instrumental sensitivity over the detector volume and all possible arrival neutrino directions.
A minimum of $10^6$ candidate neutrino events were processed in each of 17 logarithmic steps ranging from $10^{13}$~eV to $10^{21}$~eV separated by 0.5 decades in neutrino energy.
The all-sky water-equivalent volumetric acceptance (km$^{3}$~sr$^{-1}$) was then calculated for each energy as

\begin{equation}
V \Omega = \frac{4 \pi V_{\rm sim}}{N} \times \sum_i \left(p_{{\rm Earth},i} \times p_{{\rm detect},i} \times \frac{\rho_i}{\rho_{\rm water}}\right),
\end{equation}

\noindent with $i$ iterating over neutrino events.
$p_{{\rm detect},i}$ denotes probability that the electric field at station results in a trigger (see~\ref{sec:threshold}).
$\rho_i$ is the ice density at the interaction vertex and $\rho_{\rm water}$ is the density of water. 
To compute the event rate for various neutrino flux models, the sensitivity can be equivalently expressed in terms the areal acceptance (m$^{2}$~sr$^{-1}$) as

\begin{equation}
A \Omega = V \Omega / l, 
\end{equation}

\noindent where $l$ is the neutrino interaction length through water using the total interaction cross section~\cite{williams_2004_dissertation}.

\subsection{Electric field threshold for a fiducial station configuration}
\label{sec:threshold}

The final step of evaluating the neutrino acceptance is to determine which events pass the trigger.
For generality, we used a simple threshold on the electric field produced at the antennas to define a trigger criterion and we compare the acceptance of various station configurations at the trigger level.  For each case that we simulated, 
we assume the same fixed trigger rate per channel.

We assume a bandwidth of $100-800$~MHz and compute the signal electric field strength by integrating the Askaryan emission over this frequency range.
As explained in detail below, an electric field trigger threshold of 0.15 mV/m per antenna is equivalent to a $2.3\sigma$ trigger threshold for ideal dipole antennas with the assumed
bandwidth.

The root mean square voltage received in a system ($\mathrm{V}_\mathrm{RMS}$) for thermal noise is

\begin{equation}
\mathrm{V}_\mathrm{RMS}=\sqrt{k_\mathrm{B}TBZ},
\end{equation}

\noindent where $T$ is the total system temperature, $B$ is the bandwidth of the system, $Z$ is the impedance of the system, and $k_\mathrm{B}$ is Boltzmann's constant.  
Assuming a system temperature of 70~K in addition to the temperature of the ice (250~K), a bandwidth of 700~MHz, and a 50$\Omega$ system impedance, we calculate that $\mathrm{V}_\mathrm{RMS}=12.6\mu\mathrm{V}$.  
This thermal noise contribution sets the threshold of the system.

The electric field threshold is given by

\begin{equation}
E=\mathrm{V}\sqrt{\frac{Z_{0}}{A_{\mathrm{eff}} Z}},
\end{equation}

\noindent where V is the voltage received, $A_{\mathrm{eff}}$ is the effective area of the antenna, and $Z_{0}$ is the impedance of free space~\cite{balanis,kraus}.   
$A_{\mathrm{eff}}$ is defined as

\begin{equation}
A_{\mathrm{eff}}=\frac{\lambda^2 G}{4 \pi},
\end{equation}
where $G$ is the antenna gain, and $\lambda$ is the wavelength.  For a $50\Omega$ system, the electric field threshold is therefore given by

\begin{equation}
E=\mathrm{V}\frac{9.73}{\lambda\sqrt{G}}.
\end{equation}

Assuming a perfect dipole antenna with a gain of 2.15~dBi and a fast impulse response, a reference frequency of 200~MHz, and a trigger threshold of $2.3\sigma$ above thermal noise, we calculate an electric field threshold per trigger channel of 0.15~mV/m, which is the same as the per-antenna threshold for the 16-antenna station configuration without phasing.

The final acceptance for a detector must also include the efficiency of successfully reconstructing events and discriminating signal from thermal noise and anthropogenic backgrounds.
Previous radio detection experiments have achieved high analysis efficiencies relative to the trigger-level efficiency, e.g., a 96\% reconstruction efficiency for ANITA~\cite{anita2}.
Despite a lower signal-to-noise per antenna expected in a phased array configuration, the opportunity to position in-situ stations in stable, low-noise environments as well as directional rejection of backgrounds (Section \ref{sec:calc}) could result in comparably high analysis efficiencies. 

\section{Simulation diagnostics}
\label{sec:validation}

As a basic validation of the simulation output and to develop an intuition for the instrument response, we compare the distributions of several triggered event properties for different station configurations.  We investigate here two different energies (1~PeV and
1000~PeV) and the three different experimental configurations we have discussed in this paper (16-channels not phased, a 16-antenna phased array, and a 400-antenna phased array).

\begin{figure}[h]
      \begin{center}
        \includegraphics[width=7cm]{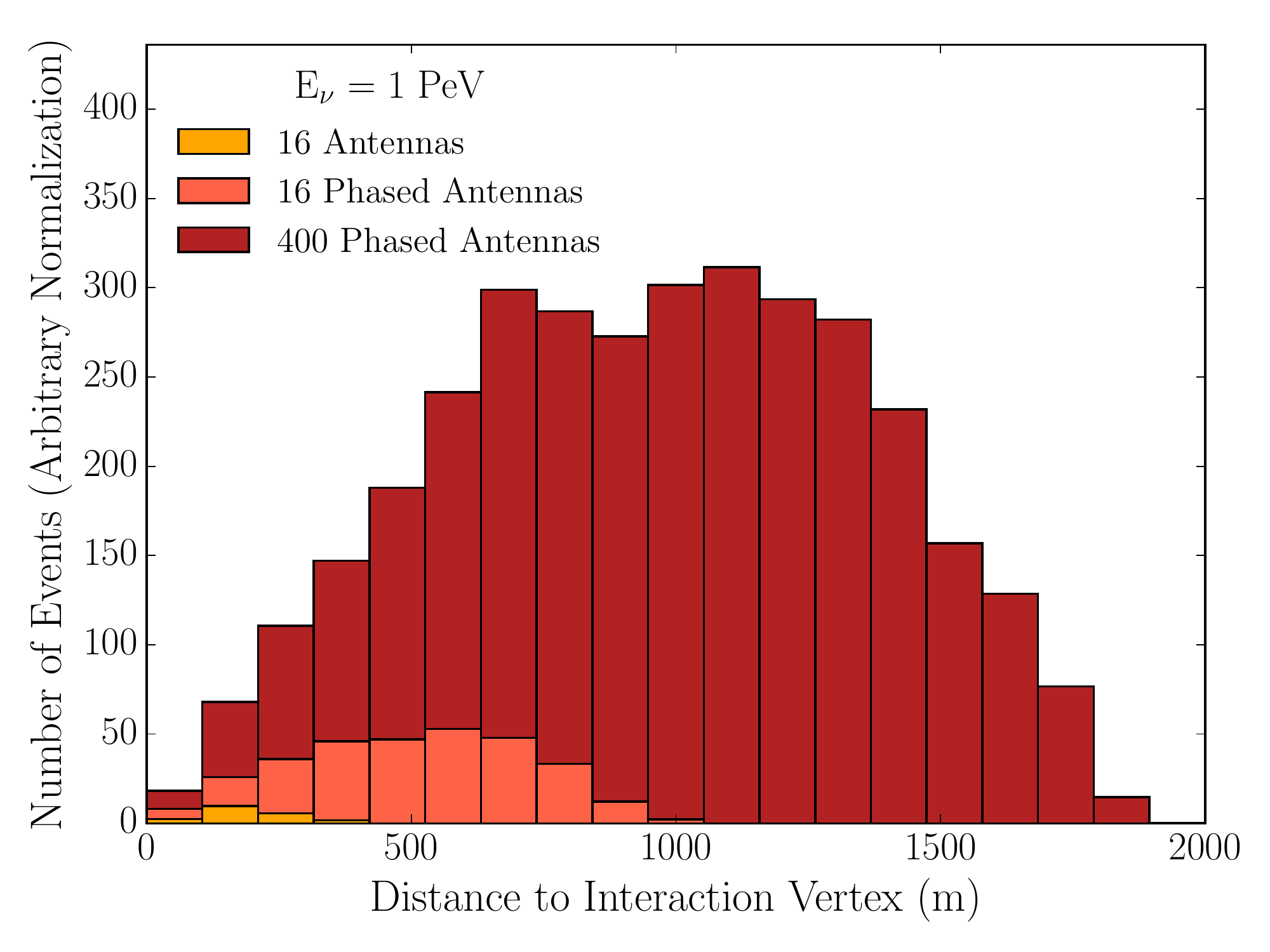}
        \includegraphics[width=7cm]{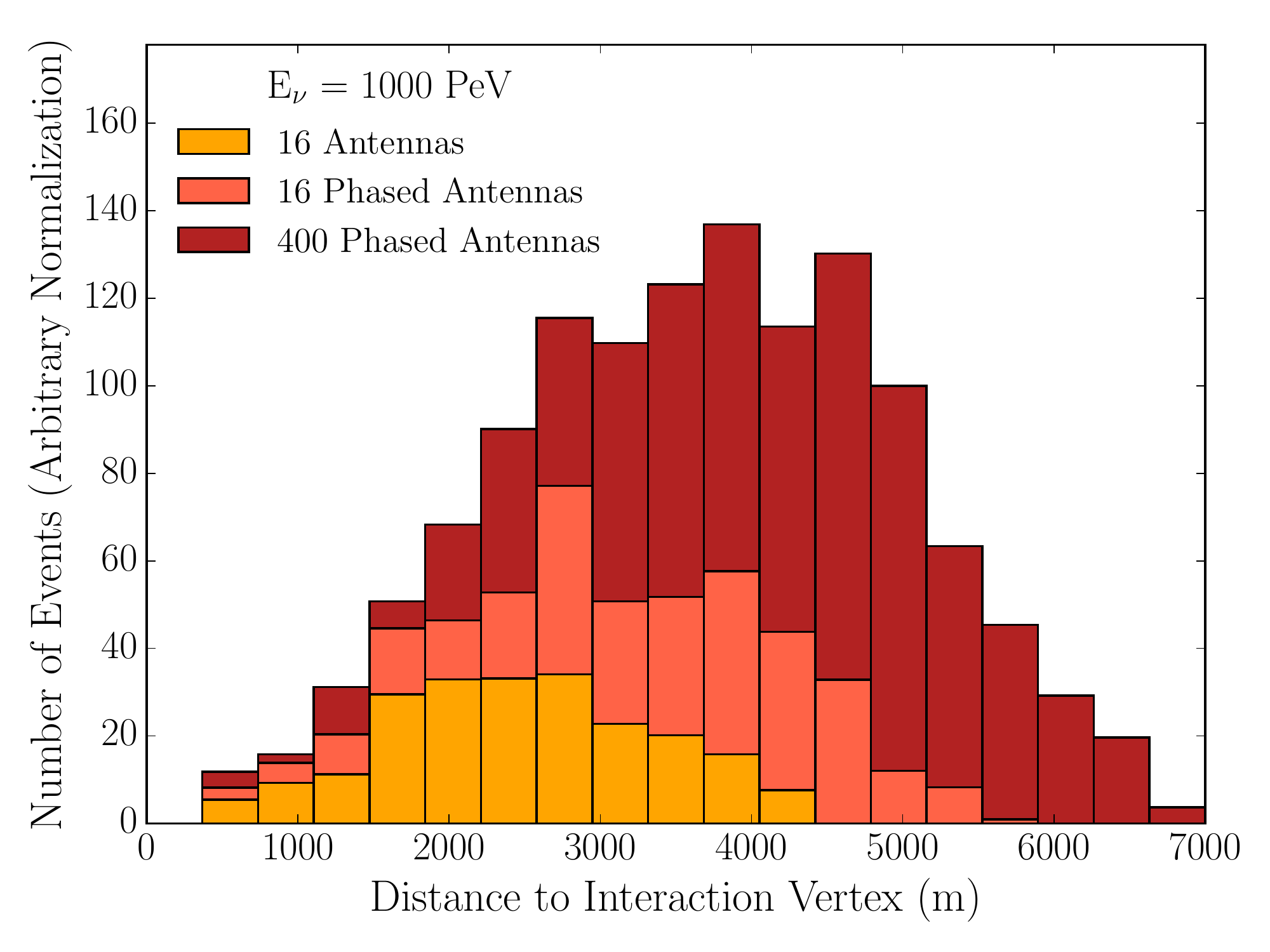}
    \end{center}
    \caption{The distance from the antennas to the interaction vertex for passing events
      at two different energies (1~PeV
      and 1000~PeV) and the three different experimental configurations we have discussed
      (16-channels not phased, a 16-antenna phased array, and a 400-antenna phased array). 
      Lower energy events occur closer to the detector, and a phased system allows more 
      distant events to be observed.  Note that the axis ranges are different between the two panels.} 
    \label{fig:distance}
  \end{figure}

Figure~\ref{fig:distance} shows the distribution of the distance from the antennas to the interaction vertex as measured along the radio signal path for each event that satisfies the trigger requirements.  
As expected, lower energy events occur closer to the detector and 
station configurations with lower thresholds are capable of triggering on more distant events.

\begin{figure}[h]
      \begin{center}
        \includegraphics[width=7cm]{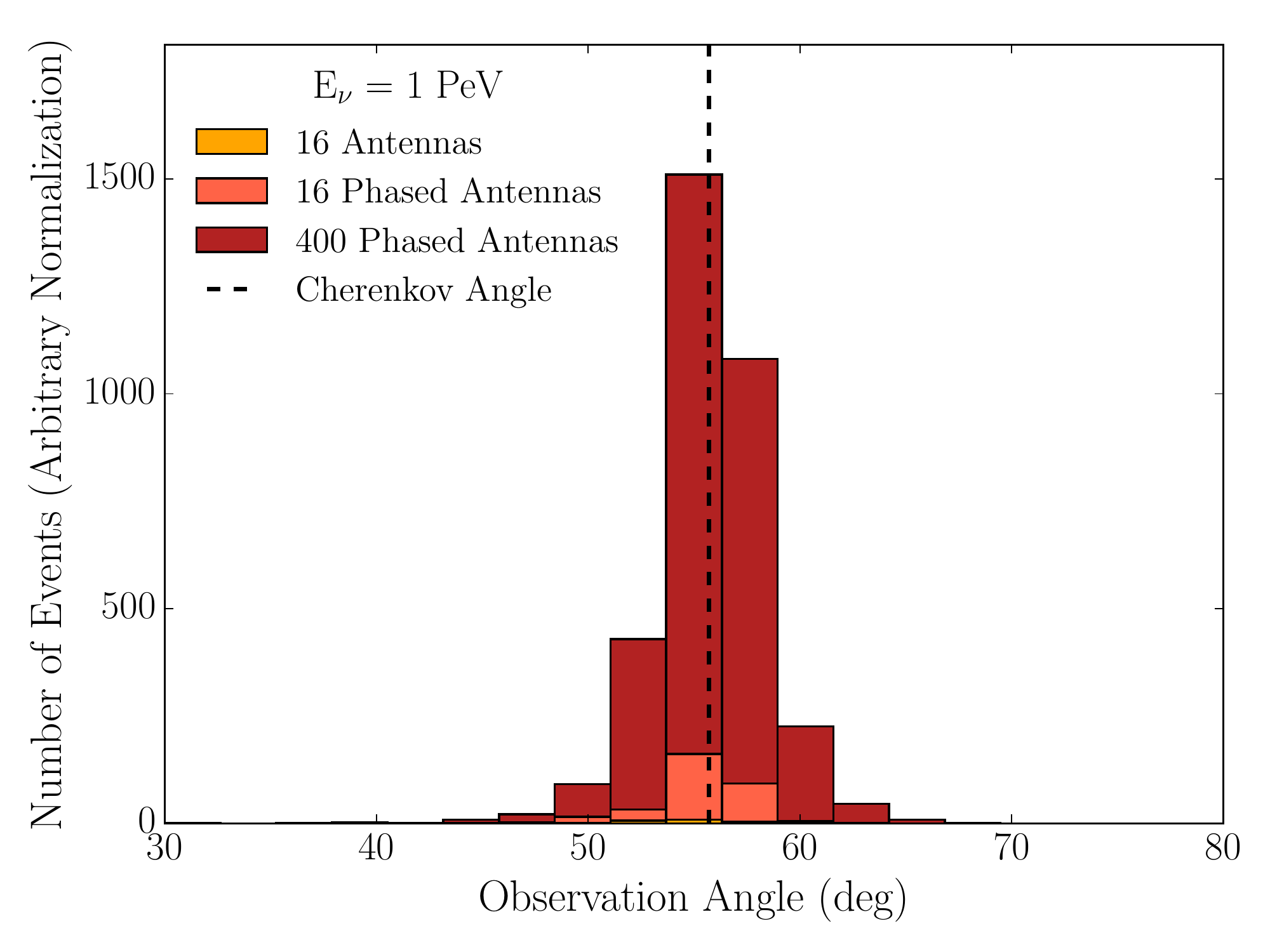}
        \includegraphics[width=7cm]{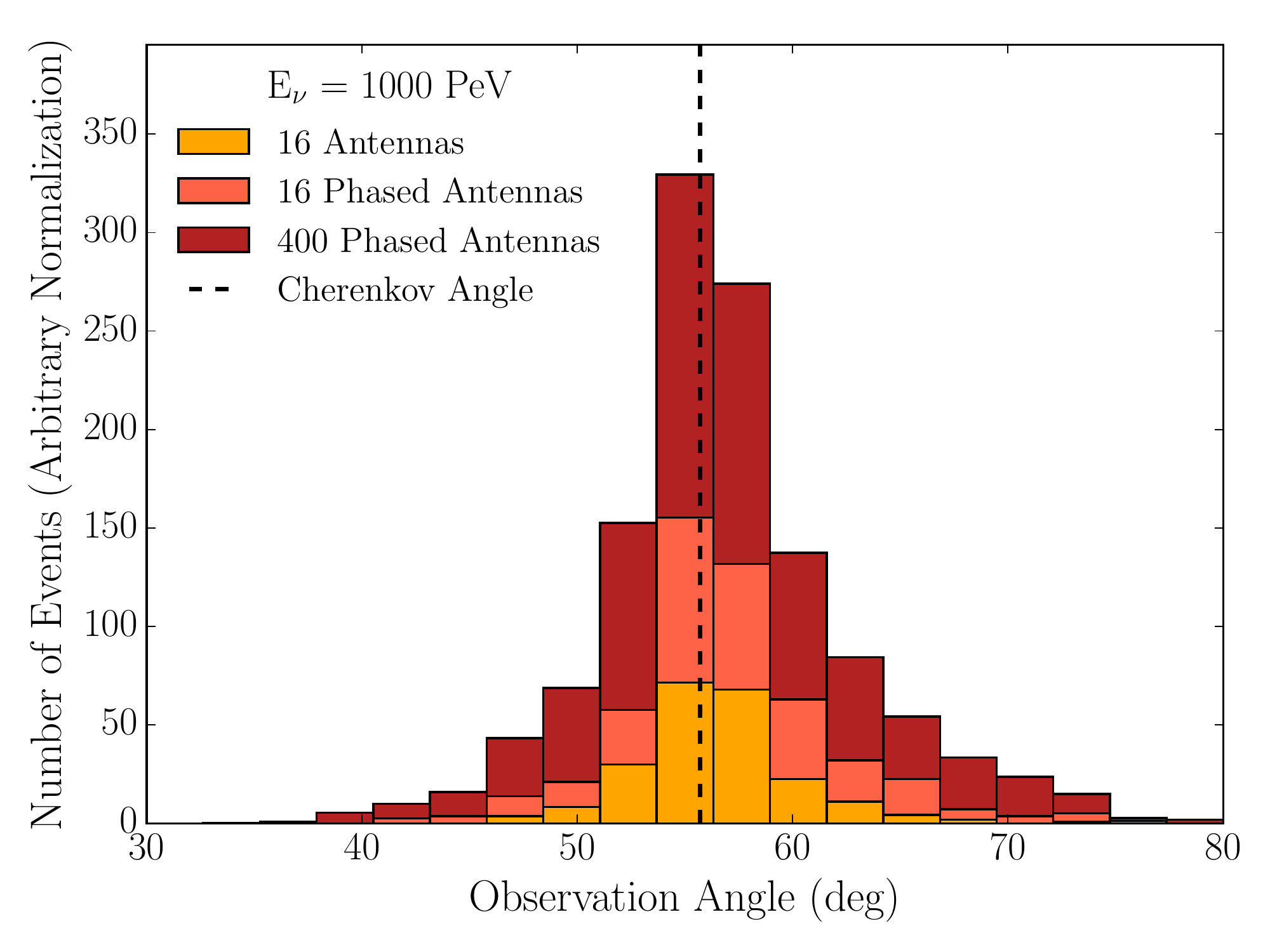}
    \end{center}
    \caption{The angle of observation of the radio signal compared to the incident neutrino propagation direction 
      for passing events at two different energies (1~PeV and 1000~PeV) and the three different experimental configurations we 
      have discussed 
      (16-channels not phased, a 16-antenna phased array, and a 400-antenna phased array).  The Cherenkov
      angle is shown with a dashed line for reference.  Higher energy events are seen farther from
      the Cherenkov angle, and a phased system allows events to be observed farther from the Cherenkov angle.} 
    \label{fig:viewangle}
  \end{figure}

The angle at which the detector views events relative to their neutrino propagation direction is shown in Figure~\ref{fig:viewangle}.
The Cherenkov angle is shown for reference as a dashed line.
As expected, higher energy events are seen farther from the Cherenkov angle.  Configurations with lower threshold (e.g., phased arrays) also allow for events to be seen farther from the Cherenkov angle where the emission is weaker.


\begin{figure}[h]
      \begin{center}
        \includegraphics[width=7cm]{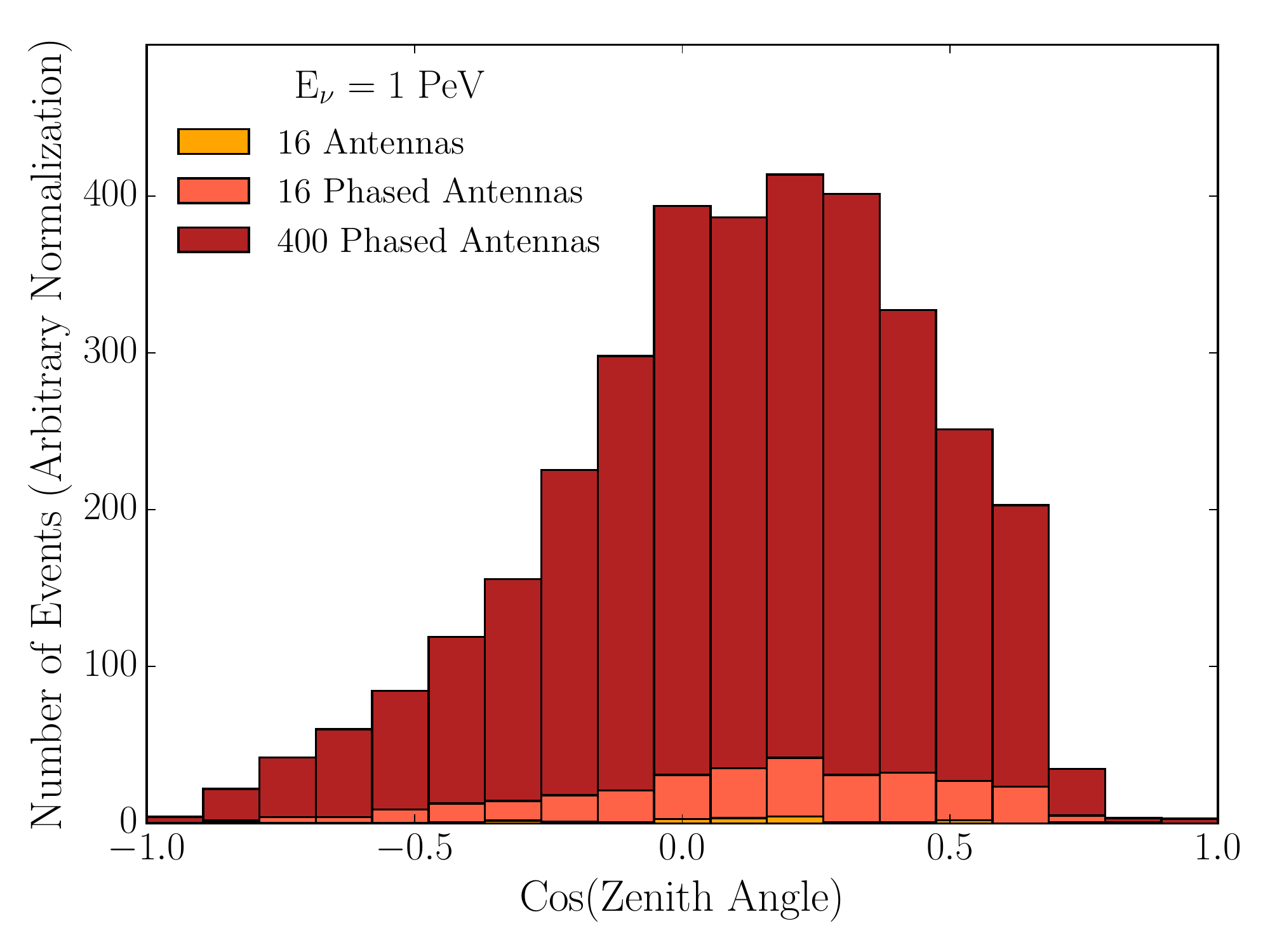}
        \includegraphics[width=7cm]{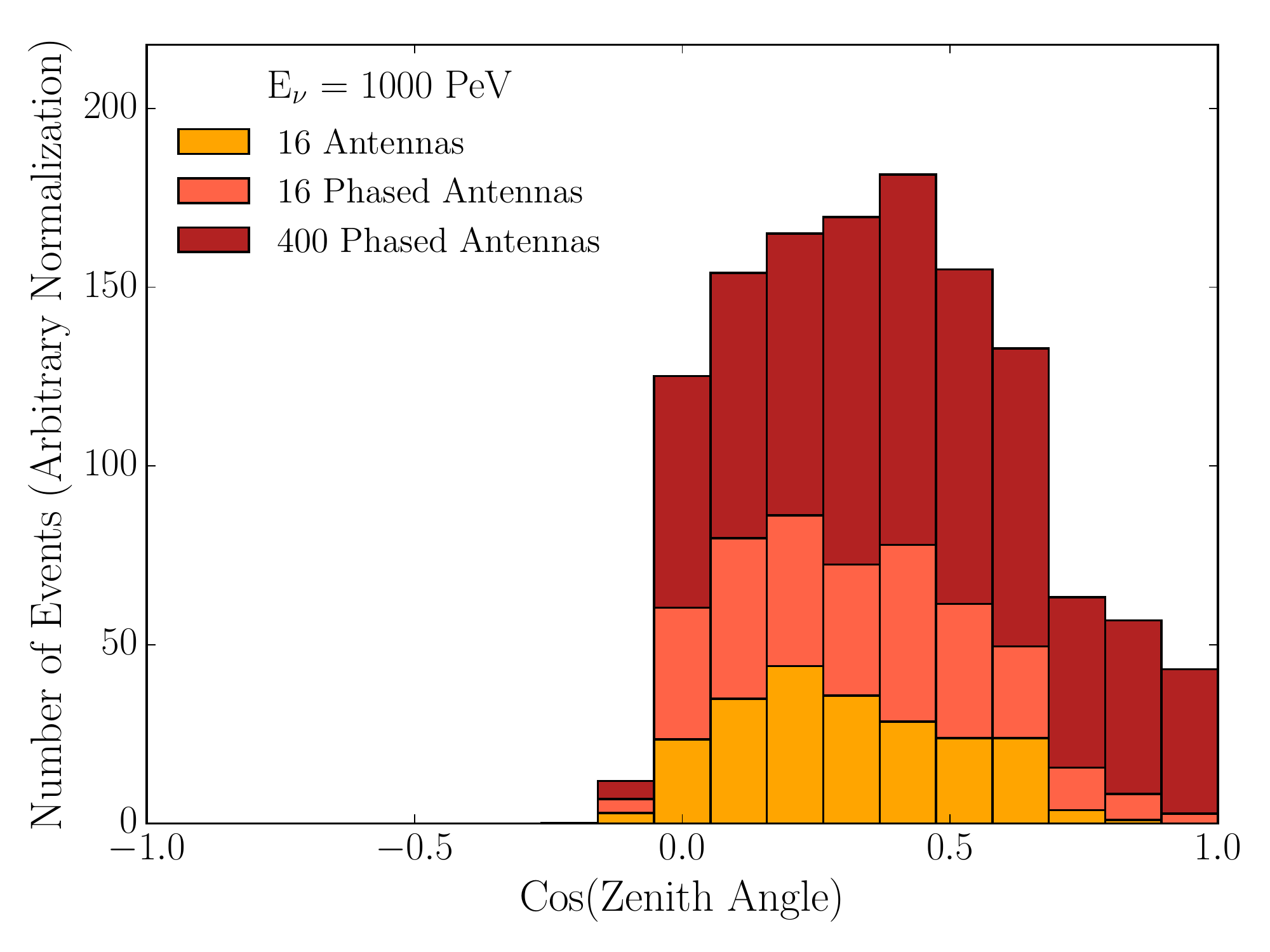}
    \end{center}
    \caption{The zenith angle of the incident neutrino direction 
      for passing events at two different energies (1~PeV
      and 1000~PeV) and the three different experimental configurations we have discussed
      (16-channels not phased, a 16-antenna phased array, and a 400-antenna phased array).  
      The arrays are most sensitive to neutrinos that are slightly down-going.} 
    \label{fig:zenith}
  \end{figure}

The distribution of incident neutrino zenith angles is shown in Figure~\ref{fig:zenith}.
The events are partitioned into bins of equal solid angle.
At 1000 PeV, the acceptance is dominated by events with arrival directions above the horizon due to effective screening by the Earth for up-going neutrinos.
As expected, the zenith angle distribution is broader for 1 PeV neutrinos, for which the interaction length is comparable to the size of the Earth.

\bibliographystyle{JHEP}
\bibliography{paper}

\end{document}